\documentclass[12pt]{article}

\def\a{\alpha}

\def\b{\beta}

\def\S{\Sigma}

\def\f{\phi}
\def\F{\Phi}

\def\D{\Delta}
\def\e{\epsilon}

\def\G{\Gamma}
\def\g{\gamma}
\def\o{\omega}
\def\O{\Omega}

\def\d{\delta}

\def\L{\Lambda}

\def\t{\tau}

\def\det{\textrm{det}}

\def\cd{{\cal D}}
\def\cc{{\cal C}}

\def\pr{\partial}
\def\to{\rightarrow}

\newcommand{\be}{\begin{equation}}
\newcommand{\ee}{\end{equation}}
\newcommand{\bea}{\begin{eqnarray}}
\newcommand{\eea}{\end{eqnarray}}

\begin{document}

\begin{titlepage}

\bigskip
\bigskip
\bigskip
\bigskip

\begin{center}

{\bf{\Large Quantum gravity vacuum and invariants of embedded spin
networks}}

\end{center}
\bigskip
\begin{center}
 A. Mikovi\'c \footnote{E-mail address: amikovic@ulusofona.pt}
\end{center}
\begin{center}
Departamento de Matem\'atica e Ci\^{e}ncias de Computac\~ao,
Universidade Lus\'ofona, Av. do Campo Grande, 376, 1749-024,
Lisboa, Portugal
\end{center}

\normalsize

\bigskip
\bigskip
\begin{center}
                        {\bf Abstract}
\end{center}
We show that the path integral for the three-dimensional SU(2) BF
theory with a Wilson loop or a spin network function inserted can
be understood as the Rovelli-Smolin loop transform of a
wavefunction in the Ashtekar connection representation, where the
wavefunction satisfies the constraints of quantum general
relativity with zero cosmological constant. This wavefunction is
given as a product of the delta functions of the SU(2) field
strength and therefore it can be naturally associated to a flat
connection spacetime. The loop transform can be defined rigorously
via the quantum SU(2) group, as a spin foam state sum model, so
that one obtains invariants of spin networks embedded in a
three-manifold. These invariants define a flat connection vacuum
state in the q-deformed spin network basis. We then propose a
modification of this construction in order to obtain a vacuum
state corresponding to the flat metric spacetime.
\end{titlepage}
\newpage

\section{Introduction}

In a recent review of loop quantum gravity \cite{Smo}, Smolin has
given a comprehensive description of a vacuum state for a quantum
general relativity (GR) theory with a positive cosmological
constant $\L$. This state can be defined in the loop
representation as a linear combination of the q-deformed spin
network states, where the coefficients are given by the loop
transforms of the Kodama wavefunction. These coefficients can be
understood as the invariants of embedded spin networks in the
spatial manifold $\S$ determined by the path integral of the
$SU(2)$ Chern-Simons (CS) theory. These invariants are
generalizations of the Jones polynomial \cite{W}, and can be
expressed as the Kaufman bracket invariants, i.e. as evaluations
of the q-deformed spin networks \cite{K,KL}, or more generally as
the Reshetikhin-Turaev (RT) invariants \cite{RT}.

However, there exists another set of spin network invariants for
the three-dimensional manifolds, which are also determined by a
path integral over the connections, but instead of the CS action
one uses the BF theory action \cite{PfO,Pf,Oe}. The purpose of
this paper is to demonstrate that these new invariants can be
understood as the loop transforms of a flat connection vacuum
wavefunction which corresponds to quantum GR with zero
cosmological constant. This vacuum wavefunction is given by a
product of the delta functions of $F$, where $F$ is an $SU(2)$
two-form field strength, and therefore that wavefunction is
annihilated by the GR constraints in the Ashtekar connection
representation. One can then naturally associate this wavefunction
to a classical vacuum solution with $F=0$. There are many such
solutions, and we will be interested in constructing a quantum
state describing the flat spacetime metric solution. Our approach
is an extension of a partially reduced phase space quantization
approach of Baez \cite{bafc}, where he imposes classically the
$F=0$ constraint and then considers the wavefunctions on the
moduli space of flat connections.

In section 2 we show how the BF theory path integral for a spin
network can be represented as the Rovelli-Smolin loop transform of
a flat connection vacuum wavefunction in the Ashtekar connection
representation. We then show how the constraints of quantum GR for
zero cosmological constant annihilate this vacuum wavefunction,
and how to construct the corresponding vacuum state in the spin
network basis. In section 3 we define the BF theory spin network
path integrals as the spin foam model state sum invariants for the
quantum $SU(2)$ group. In section 4 we present our conclusions and
discuss our results, including a discussion on the relationship
between the $F=0$ classical solutions and the flat spacetime
metric solution. We then propose a modification of our
construction in order to obtain a flat spacetime metric vacuum
state.

\section{Ashtekar formulation}

The Ashtekar formulation of GR \cite{A}, can be understood as a
special choice of variables in the canonical formulation of GR
based on the triad one forms $e^a = e_i^a dx^i$, where $a=1,2,3$,
and $x^i$ are the coordinates on the spatial three-manifold $\S$.
The three-metric is given by $h_{ij}=e_i^a e_{ja}$ and the
spacetime manifold is given by $\S\times{\bf R}$. Let $e^i_a$ be
the inverse triad components, and let $p^a_i$ be the corresponding
canonically conjugate momenta. Then the Ashtekar variables can be
defined as \be A_i^a = \o_i^a (e) + i {p_i^{a}\over \sqrt{h}}
\quad,\quad E^i_a = \sqrt{h}\, e^i_a \quad,\label{ashv}\ee where
the coefficient $i=\sqrt{-1}$, $h=\det(h_{ij})$ and the $\o(e)$ is
the torsion-free $SO(3)$ spin connection. Hence the new connection
$A$ is complex, and the advantage of this unusual choice is that
the Hamiltonian constraint becomes polynomial in the new canonical
variables $(-iA,E)$, i.e. \be C_H = \e_{abc}\left( E^{ia} E^{jb}
F^{c}_{ij} + \L\e_{ijk} E^{ia} E^{jb} E^{kc} \right)
\quad,\label{hamc} \ee where $F$ is the curvature two form
associated to the $SU(2)$ connection $A$ and $\L$ is the
four-dimensional cosmological constant. The Gauss and the
diffeomorphism constraints are also polynomial in the new
variables \be C_a = \pr_i E^i_a + \e_{abc}A_i^b E^{ic} \quad,\quad
C_i = F_{ij}^a E^j_a \quad.\label{gdc}\ee

Because of this simplification, one can adopt a Dirac quantization
strategy where the pair $(A,E)$ is represented by the operators
\be \hat{A}_i^a = A_i^a \quad,\quad\hat{E}^i_a = {\d\over \d
A_i^a}\quad,\label{hor}\ee acting on the wavefunction(al)
$\Psi(A)$, such that \be \hat C_a \Psi = 0 \quad,\quad \hat C_i
\Psi = 0 \quad,\quad \hat C_H \Psi = 0 \quad,\ee where the $\hat
C$ operators are obtained by substituting (\ref{hor}) in the
classical expressions in a chosen order. The $SU(2)$ connection
$A$ is complex, and the reality properties of the expressions
(\ref{ashv}) require that the wavefunction $\Psi(A)$ should be a
holomorphic functional of $A$, i.e. independent of $\bar A$
\cite{Ko2}. This holomorphy also implies that in order to
construct a wavefunction of the triads $e$, it is sufficient to
consider only the real $A$'s \cite{Ko2}.

Hence by constructing a gauge invariant and a diffeomorphism
invariant functional of the $SU(2)$ connection $A$ defined on the
three-manifold $\S$, one will solve the Gauss and the
diffeomorphism constraints. Therefore the problem is reduced to
analyzing the action of the Hamiltonian constraint operator $\hat
C_H$. For example, the Kodama wavefunction  \be \Psi_K [A] = \exp
\left( {1\over 2 \L}\int_\S Tr\,(A\wedge dA + {2\over 3}A\wedge
A\wedge A )\right) = e^{kS_{CS}[A]}\quad,\ee which is an exponent
of the Chern-Simons action, is gauge and diffeomorphism invariant
functional and satisfies the Hamiltonian constraint for a non-zero
cosmological constant \cite{Ko}.

A more advanced strategy of quantization is to use the
Rovelli-Smolin loop representation \cite{RSlr}, i.e. if we think
of $\Psi (A)$ as $\langle A|\Psi\rangle$, then one can introduce
the loop states $|\g\rangle$ associated to the Wilson loop
functionals $W[\g,A]=Tr\,P\exp\int_\g A$, and define the loop
representation as \be \Psi(\g)=\langle\g|\Psi\rangle =\int \cd A
\langle\g |A\rangle\langle A|\Psi\rangle = \int \cd A\,
W[\g,A]\,\Psi(A)\quad.\label{ltr}\ee If $\Psi(A)$ is a solution of
the gauge and the diffeomorphism constraints (\ref{gdc}), then
$\Psi(\g)$ will be a three-manifold invariant of the loop $\g$.
More generally, one can replace the $\g$ with a labelled graph
$\G$, and hence obtain an invariant of the embedded spin network.
In the case of the Kodama wavefunction, one can argue that
$\Psi(\G)$ should be the invariant given by the Kauffman bracket
\cite{MS}\footnote{This argument is straightforward in the case of
the Euclidian GR, when the Ashtekar connection is real and the
Kodama wavefunction is $\exp(ikS_{CS})$. In the Lorentzian case,
$A$ is complex, and the Kodama wavefunction is $\exp(kS_{CS})$, so
that a factor of $i$ is missing. There is no rescaling $A\to cA$,
where $c$ is a complex number, which will give that factor, so
that the argument is that a Wick rotation $k\to ik$ of the
Euclidian result has to be done.}. This is a consequence of the
fact that the CS action defines a topological theory on $\S$, and
the path integral (\ref{ltr}) defines an observable of the theory
\cite{W}.

However, there exists another topological theory in three
dimensions, closely related to the CS theory, namely the BF theory
(for a review and references see \cite{bar}). It is also a theory
of flat connections, but its phase space is bigger, because it
includes a one form $B$ field. The BF $SU(2)$ theory can be
considered as the three-dimensional Euclidian GR with a zero
cosmological constant \cite{bar}. The partition function of this
theory \be Z_{BF} = \int \cd A \,\cd B\, e^{i\int Tr\,(B\wedge
F)}\quad,\label{bfpi}\ee can be defined via simplical
decomposition of the manifold as a spin foam state sum
\cite{boul,fksf,bar}. This gives the Ponzano-Regge state sum
model\cite{PR}. A topologically invariant way to regularize the
Ponzano-Regge state sum is via the $SU(2)$ quantum group, so that
one obtains the Turaev-Viro state-sum model \cite{TV}.

Given these properties, one is led to consider the following
quantity \be \Psi_0 (\g) = \int \cd A\,\cd B\, W[\g,A]\,e^{i\int
Tr\,(B\wedge F)}\quad,\label{bfwl}\ee i.e. the BF theory analog of
the CS Wilson loop expectation value. This quantity should be
gauge and diffeomorphism invariant, and it can be defined via
simplical decomposition of $\S$ as a spin foam state sum
\cite{PfO,Pf,Oe}. This state sum can be then regularized in a
topologically invariant way via the quantum $SU(2)$ group.

The form of the path-integral $\Psi_0$ is such that after
performing the $B$ integration we obtain \be \Psi_0 (\g) = \int
\cd A\, W[\g,A]\, \d (F)\quad,\label{lt}\ee where \be\d(F)=\int
\cd B e^{i\int Tr\,(B\wedge F)} =
\prod_{x\in\S}\prod_{i,j}\d(e^{F_{ij}(x)}) \quad.\ee These are
heuristic expressions, which have to be properly defined (see the
next section), but for the moment if we think of $\d (F)$ as a
functional of the connection $A$, then $\d (F) = \Psi_0(A)$ is
gauge invariant due to group delta function identity $\d (g e^F
g^{-1}) = \d (e^F)$. Furthermore we have \be \hat C_i \, \Psi_0 =
{\d\over \d A_j^a(x)}F_{ij}^a
(x)\prod_{y\in\S}\prod_{k,l}\d(e^{F_{kl}(y)})= 0 \quad,\ee and \be
\hat C_H \, \Psi_0 =\e^{abc}{\d\over \d A_i^a(x)}{\d\over \d
A_j^b(x)}F_{ijc}(x)\prod_{y\in\S}\prod_{k,l}\d(e^{F_{kl}(y)})= 0
\quad,\ee due to $F(x) \d (e^{F(x)}) =0$. Hence $\Psi_0 (A)$ is
gauge and diffeomorphism invariant functional which satisfies the
Hamiltonian constraint for the zero cosmological constant.

This heuristics implies that the invariant (\ref{bfwl}) is a loop
transform of a quantum gravity $F=0$ vacuum state $\Psi_0
(A)=\d(F)$. This vacuum state can be naturally associated to the
space of $F=0$ classical solutions of the Einstein equations.
A particular $F=0$ solution is a flat metric space when $A=\omega
(e)$. Hence understanding the properties of the $\d (F)$ state will
be a first step toward constructing a flat metric vacuum state.

One can construct a flat connection vacuum state in the
diffeomorphism invariant basis of spin network states $|\G\rangle$
\cite{RSsn}, as \be |0\rangle = \sum_\G
|\G\rangle\langle\G|0\rangle \quad, \ee where
 \be \langle\G|0\rangle = \Psi_0 (\G) =\int\cd A\, W[\G,A]\, \d (F)
= \int \cd A\,\cd B\,W[\G,A]\,e^{i\int Tr\,( B\wedge F)}\quad,
\label{snin}\ee is a spin
network invariant to be defined, and $W[\G,A]$ is the spin network
generalization of the Wilson loop functional, i.e. a product of
the edge holonomies contracted by the vertex intertwiners.

\section{Spin-foam state sum representation}

Let us now write the spin foam state sum representation of the
spin network invariant (\ref{snin}) in order to better understand
its properties. This path integral can be defined via the
discretization procedure provided by a simplical decomposition of
$\S$. Let $\cc$ be a simplical complex associated to $\S$, and let
$\cc^*$ be the dual complex. We denote by $\t$, $\D$ and $\e$ a
tetrahedron, a triangle and an edge of $\cc$ respectively, and by
$v$, $l$ and $f$ the corresponding dual cells, i.e. a vertex, a
link and a face. Let the $B$ one-form be piece-wise constant in
the 3-polytopes formed by the pairs $(\e,f )$, where $f=\e^*$,
such that the only non-zero spatial components of $B$ are those
parallel to the vectors $\e$\footnote{This is an approach to the
discretization introduced in \cite{mym}. It is a more convenient
approach for treating non-linear actions then the usual approach
where the $B$ field is taken to be a distribution concentrated
along the edges $\e$ \cite{fksf,bar}.}. Then \be \int_\S
Tr\,(B\wedge F) = \sum_{\e\in\cc}Tr\,( B_\e F_f )
 \quad,\ee where \be B_\e = \int_\e B \quad,\quad F_f = \int_f F
\label{simpa}\quad.\ee

Given the variables (\ref{simpa}), we can define the path integral
(\ref{bfpi}) as \be Z = \int \prod_{l} dA_l \prod_{\e} dB_\e \,
e^{i\sum_{\e}Tr\,(B_{\e} F_f )} \quad,\ee where $A_l = \int_l A$.
We will fix the measures of integration $dA$ and $dB$ such that
\be Z = \int \prod_l dg_l \prod_f \d \left(\prod_{l\in
\partial f} g_{l}\right) \quad,\label{gtz}\ee where $g_l =
e^{A_l}$. Note that (\ref{gtz}) is a discretized form of the path
integral $\int\cd A \d (F)$ and therefore $\prod_f \d
\left(\prod_{l\in
\partial f} g_{l}\right) = \prod_f \d \left( e^{F_f}\right)$ can
be considered as a discretized form of the wavefunction $\d(F)$.

By using the group theory formulas \be \d (g) = \sum_{j}\textrm{
dim}\,j \,\chi_j (g)= \sum_{2j=0}^{\infty} (2j+1) \,\chi_j
(g)\quad, \label{delta}\ee and \be \int_{SU(2)} \,dg
\,D^{(j_1)}_{\a_1\b_1} (g) D^{(j_2)}_{\a_2\b_2}
(g)D^{(j_3)}_{\a_3\b_3} (g)  = C^{{j_1}{j_2}{j_3}}_{\a_1 \a_2
\a_3} \left(C^{{j_1}{j_2}{j_3}}_{\b_1 \b_2
\b_3}\right)^*\quad,\label{threej}\ee where $D^{(j)}$ is the
representation matrix in the representation $j$, $\chi_j$ is the
corresponding trace and $C$'s are the Clebsh-Gordan coefficients,
or the $3j$ intertwiners, as well as the associated graphical
calculus \cite{fksf,bar}, one arrives at \be Z = \sum_{j_f
}\prod_f \textrm{ dim}\,j_f \prod_v A_{tet}
(j_{f_1(v)},...,j_{f_6(v)})\quad. \label{3dsf}\ee $A_{tet}$ is the
tetrahedron spin network amplitude, given by a contraction of four
$3j$ tensors\footnote{For a tensorial approach to spin networks
see \cite{msf}.}, which is also known as the $(6j)$ simbol. The
terms in the
sum (\ref{3dsf}) can be interpreted as the amplitudes for a colored
two-complex associated to $\cc^*$, which is called the spin-foam
\cite{bar}. The Ponzano-Regge state sum formula can be obtained by
by relabelling the sum (\ref{3dsf}) as a sum over the  irreducible
representations (irreps) which are associated to the simplices
dual to the spin foam, i.e. the edges and the tetrahedrons of the
manifold triangulation\be Z = \sum_{j_\e }\prod_\e \textrm{
dim}\,j_\e \prod_\t A_{tet} (j_{\e_1(\t)},...,j_{\e_6(\t)})=
\sum_{j_\e }\prod_\e (2j_\e + 1) \prod_\t (6j)(\t) \quad.
\label{prss}\ee

The expression (\ref{prss}) is not finite, because the infinite
sum over the $SU(2)$ irreps $j$ divereges. Formally, the sum
(\ref{prss}) is topologically invariant, because it is invariant
under the three-dimensional Pachner moves. It is possible to
regularize this sum while preserving the topological invariance,
by replacing the $SU(2)$ irreps with the irreps of the quantum
group $SU_q (2)$ for $q=e^{i\pi /n}$, where $n$ is an integer
greater than 2 \cite{TV}. Then for each $n$ there are finitely
many irreps, up to $j={n-2\over 2}$. Each face (edge) amplitude
$\textrm{dim}\,j$ is replaced by $\textrm{dim}_q \,j$, while each
vertex (tetrahedron) amplitude $(6j)$ is replaced by $(6j)_q$. In
this way one obtains the Turaeev-Viro state sum invariant \be
Z_{TV} = d_k^{-2V}\sum_{j_\e }\prod_\e \textrm{dim}_q j_\e
\prod_\t (6j)_q (\t) \quad, \label{tvss}\ee where
$d_k^2 =\sum_{j=0}^{k/2}(\dim_q j)^2$, $V$ is the number of
tetrahedra and the $(6j)_q$ symbol is
appropriately normalized.

In the case of the path integral (\ref{snin}), we will use the
formula \be \Psi_0 [\G] = \int \prod_l dg_l \prod_f \d
\left(\prod_{l\in
\partial f} g_{l}\right)W_\G (g_{l^\prime},s_{l^\prime}, i_{v^\prime})
 \quad,\label{snpi}\ee
where $W_\G$ is the simplical version of the spin network function
for the spin network $\G$. This embedded spin network is specified
by labelling a subset $\{l^\prime\}$ of the dual edges by the
irreps $\{s_{l^\prime}\}$, as well as by labelling the
corresponding set of vertices $\{v^\prime\}$ with the appropriate
intertwiners $i_{v^\prime}$. One can associate more than one irrep
to a given link, in order to obtain the spin nets of the valence
higher than four. By using (\ref{delta}), (\ref{threej}) as well
as \be \int_{SU(2)} \,dg \,D^{(j_1)}_{\a_1\b_1} (g)\cdots
D^{(j_k)}_{\a_n\b_n} (g)  =
\sum_{\iota}C^{{j_1}\cdots{j_n}(\iota)}_{\a_1 \cdots \a_n}
\left(C^{{j_1}\cdots{j_n}(\iota)}_{\b_1 \cdots
\b_n}\right)^*\quad,\quad n > 3 \quad,\ee for the edges which
carry the irreps of the spin network, where $C^{(\iota)}$ are the
intertwiners for $n$ irreps, one obtains \be \Psi_0 (\G) =
\sum_{j_f ,\iota_{l^\prime} }\prod_f \textrm{dim}\,j_f \prod_{v\in
V\setminus V_\G} (6j)(v)\prod_{v^\prime \in V_\G}\tilde
A_{tet}(j_{f(v^\prime)},s_{l^\prime(v^\prime
)},i_{v^\prime},\iota_{l^\prime(v^\prime )})\quad. \label{snss}
\ee
The amplitude $\tilde A_{tet}$ is the evaluation for the spin network
consisting of the tetrahedral graph labelled by the $j_f$ irreps,
with an additional vertex labelled by the intertwiner $i_{v^\prime}$,
which is connected by the edges labelled by the irreps $s$ to the
tetrahedral graph vertices \cite{PfO,Pf}. This
type of the spin network amplitudes also appears in the case of
the spin foam models proposed for matter \cite{msf}, and the spin
foam amplitude (\ref{snss}) is a special case of a more general
spin foam amplitude proposed for an embedded open spin network
\cite{msf,mct}.

As in the case of the partition function, the expression
(\ref{snss}) is a divergent sum. One can then regularize it by
using the $SU_q(2)$ irreps. The reason why this procedure works,
is that the representations of $SU(2)$ and the representations of
$SU_q(2)$ form the same algebraical structure, i.e. the tensor
category (see \cite{Oe} for a review and references). The
difference is that the $SU_q (2)$ category is a more general type
of a tensor category, i.e. a ribbon tensor category, whose
algebraical properties are defined by the invariance properties of
ribbon tangles under the isotopies in ${\bf R}^3$ or $S^3$. The $3j$
and the $6j$ symbols, as well as the other spin network amplitudes
can be interpreted as morphisms in this category, i.e. the maps
among the objects of the category and the compositions of these
maps. These maps can be represented by the open and the closed spin
network diagrams. The relations among the intertwiner morphisms
which imply the topological invariance are preserved and
well-defined in the quantum group case. Hence the quantum group
version of the embedded spin network state sum (\ref{snss}) should
be a topological invariant.

Therefore a well-defined invariant for $\G$ will be given by the
quantum group version of the spin foam amplitude (\ref{snss}) \be
\Psi_0 (\G) = d_k^{-2V}\sum_{j_f ,\iota_{l^\prime} }\prod_f
\textrm{dim}_q
\,j_f \prod_{v\in V\setminus V_\G} (6j)_q (v)\prod_{v^\prime \in
V_\G}\tilde A_{tet}^{(q)}(j_{f(v^\prime)},s_{l^\prime(v^\prime
)},i_{v^\prime},\iota_{l^\prime(v^\prime )})\quad.
\label{qsnss}\ee The face and the vertex amplitudes are now given
by the evaluations of the corresponding q-deformed spin networks.
Nontrivial dual edge (triangle) amplitudes may appear, although
these can be reabsorbed into normalization factors of the vertex
amplitudes. In order to determine these normalization factors it
will be necessary to check the corresponding identities for the
simplex amplitudes associated to the invariance under the motions of
the graph in the manifold.

\section{Discussion and Conclusions}

In order for (\ref{qsnss}) to be well-defined, the spin network
$\G$ should be framed, i.e. $\G$ should be a ribbon graph spin
network. The invariant (\ref{qsnss}) could be interpreted as a
coefficient of a state $|0_q \rangle$ in the diffeomorphism
invariant basis of q-deformed spin network states $|\G_q\rangle$
\cite{Smo}, i.e. \be |0_q\rangle = \sum_{\G_q} |\G_q\rangle
\langle\G_q|0_q\rangle = \sum_{\G_q} |\G_q\rangle \Psi_0
(\G)\quad. \label{qvac}\ee The q-deformed spin network states can
be defined via the q-deformed loop algebra \cite{MS}, and these
states have the same labels as the non-deformed ones, but the
q-deformed states have the restriction that the the maximal label
$j$ is given by $(n-2)/2$ where $q=\exp(i\pi /n)$. Since the
q-deformed spin networks are naturally associated to the quantum
GR with $\L \propto 1/n$ \cite{MS,Smo}, then we expect that the
action of the $\L=0$ Hamiltonian constraint (HC) operator on the
state (\ref{qvac}) should give a state with the basis coefficients
of order $1/n^2$ or higher. Verifying this would require the
knowledge of the matrix elements of the HC operator in the
q-deformed spin network basis.

Given that the $\L=0$ quantum GR is naturally formulated in the
non-deformed ($q=1$) spin network basis \cite{RSsn,Smo}, one would
like to construct a flat connection vacuum state in that
basis. Since $q=e^{i\pi/n}$, a natural way would be to consider a
large $n$ limit of (\ref{qvac}), i.e.
\be |0\rangle = \lim_{n\to\infty} |0_q\rangle  \quad.\label{zvac}
\ee
It would be interesting to study the action of the HC operator on the state
(\ref{zvac}), given that the matrix elements of a HC operator
corresponding to the constraint (\ref{hamc}) for $\L =0$ are known
in the non-deformed spin network basis \cite{mehc}.

Note that the state $|0\rangle$ would determine a wavefunction
$\Psi_0 (A)$ via \bea\Psi_0 (A) &=&\langle A|0\rangle =\sum_{\G}
\langle A|\G\rangle \langle \G |0\rangle=\sum_{\G}
\overline{W}[\G,A] \Psi_0 (\G)\\&=&\sum_{\G} \overline{W}_\G
(g_1,\cdots,g_L) \Psi_0 (\G) \quad,\label{vwf}\eea where $g$ are
the group elements associated to the spin network edge holonomies
$P\exp \int_\g A$. This expression defines an inverse loop
transform. It would be important to understand the mathematical
properties of this type of expression, because it could serve as a
definition of the vacuum wavefunction which was given by the
heuristic expression $\prod_{x\in \S} \d(e^{F_x})$. Related to
this, one would like to see what would be the result of the action
of the constraint operators on the expression (\ref{vwf}).

In the case of $\L\ne 0$ quantum GR, one can construct in the
$\G_q$ basis a vacuum state \be |0_K\rangle =\sum_{\G_q}
|\G_q\rangle \langle\G\rangle_K \quad, \label{dsvac}\ee whose
expansion coefficients are given by the Kauffman bracket
evaluation of the spin network $\G$ \cite{Smo}. The state
$|0_q\rangle$ is different from the state $|0_K\rangle$ because
the invariants $\Psi_0 (\G)$ and $\langle \G\rangle_K$ are not the
same. The invariant $\Psi_0 (\G)$ is more complicated, since it is
a sum of products of the Kaufmann brackets, which are the
q-deformed spin network evaluations. Also, $\Psi_0 (\G)$ is by
construction sensitive to the topology of the manifold, while
$\langle \G\rangle_K$ is essentially an invariant for a spin
network embedded in $S^3$ or $R^3$. Since $\Psi_0 (\G)$
corresponds to $\L=0$ quantum GR, this raises a question how to
generalize the formula (\ref{dsvac}) to the case of arbitrary
manifold topology.

This can be done by replacing the basis coefficients of the state
$|0_K\rangle$ by the RT invariants for the $\G$'s
$\langle\G\rangle_{RT}$ \cite{RT}. These invariants correspond to
the CS theory path integrals, whose construction is based on the
surgery representation of the manifold $\S$ and the representation
theory of $SU_q (2)$. The RT invariant for $\G$ does not have a
state-sum representation, although it is related to the TV
invariant for $\G$ \cite{Oe,fkrtw,barp}. The TV invariant is given
by the TV partition function for a manifold triangulation where
there is no summation over the edge (face) irreps belonging to the
embedded spin network. The TV and the RT invariants are related
because the TV partition function is the square of the modulus of
the RT partition function. The fact that the invariant $\Psi_0
(\G)$ has a very different form compared to the corresponding RT
or TV invariants, which are naturally associated to the $\L \ne 0$
quantum GR, is consistent with our conjecture that $\Psi_0 (\G)$
describes a state in the $\L=0$ quantum GR.

In order to interpret physically the vacuum states $|0_q\rangle$
and $|0\rangle$ we need the complex connection representation and
the requirement of the holomorphy of the corresponding
wavefunction. Although a very elegant formalism, it is difficult
to implement the reality conditions by using a scalar product. We
believe that this is not such a big problem, because the
holomorphic representation seems to capture the classical reality
conditions, at least in the case of finitely many degrees of
freedom \cite{Ko2}. Also, the implementation of the reality
conditions via the usual scalar product presupposes that the
scalar products of the physical states are finite. We believe that
this assumption is too strong, because it is based on the
interpretation of the square of the modulus of the wavefunction of
the universe as a probability, which is a meaningless concept.

Given the $|0_K\rangle$ or the $|0_q\rangle$ state, one is
wondering is it possible to define an inverse loop transforms for
them. On the technical side, since the algebra of functions on a
Hopf algebra is sufficiently well understood \cite{Oe}, it should
be straightforward to generalize the expression (\ref{vwf}) to the
case of functions on the quantum group. On the conceptual side,
the resulting connection representation would correspond to a
gauge theory based on the quantum group, whose properties are not
well understood.

The $F=0$ wavefunction for the $\L=0$ GR has a simple form in the
connection representation, i.e. $\Psi_0 (A) = \d (F)$. One can
then naturally associate a space of $F=0$ classical solutions to
this state. However, it is clear that no particular $F=0$
classical solution can be associated to it, and hence we cannot
associate this state to a flat metric vacuum because a metric
flatness condition was never imposed, even classically. An $F=0$
classical solution will be a flat metric solution when $A = \omega
(e)$ constraint is imposed.

Therefore if we want to construct a state which corresponds to a
flat metric spacetime, we would need to impose a flat-metric
condition quantum mechanically. In order to do this, let us
consider the equation \be \left(\hat A_i^a - \hat\omega_i^a
(e)\right)|\F_0\rangle = 0 \quad.\ee In the ``E" representation
this equation takes a simple form \be \left(-{\d \over \d E^i_a} -
\omega_i^a (E)\right)\F_0 (E) = 0 \quad,\ee which has a solution
\be \F_0 (E) = e^{-\O (E)} \quad,\quad \O (E) =\int_\S d^3 x
\,E^i_a E^j_b \,\pr_k {e_{lc} \over\sqrt h}\, \D_{ij}^{kl}\e^{abc}
\quad,\ee where $\D$ is a tensor made from the $\d$'s such that
\be {\d \O \over \d E^i_a (x)}= \o_i^a (e)\quad.\ee In the ``A"
representation this solution will be given by \be \F_0 (A) = \int
\cd E \,e^{i\int AE d^3 x}\, \F_0 (E) \quad,\ee and let us then
consider the following loop transform \be \langle \G |0_h \rangle
= \int \cd A \, W[\G,A]\, \d (F)\, \F_0 (A) \quad.\label{flspi}\ee

This suggests a flat metric vacuum wavefunction $\Psi_0 (A)= \d (F)
\F_0 (A)$. However, although it is annihilated by the constraint
operators,
$\Psi_0$ is not annihilated by the $\hat A -\hat\o(E)$ operator.
In order to understand this, note that
the functional integral (\ref{flspi}) can be also interpreted as a
functional integral over the flat connections $A_0$ \be \langle \G
|0_h \rangle = \int \cd A_0 \, W[\G,A_0]\, \F_0 (A_0)
\quad,\label{flcpi}\ee so that (\ref{flspi}) can be interpreted as
a loop transform in the partially reduced phase space quantization
approach of Baez \cite{bafc}. Since $\F_0 (A_0)$ is annihilated by
$\hat A - \hat\o (E)$, it would make sense to define a flat
metric vacuum state as
\be |0_h \rangle = \sum_\G |\G \rangle \langle \G
|0_h \rangle \quad.\label{flsv}\ee

Given the representation (\ref{flspi}), one could hope that the
coefficients $\langle \G |0_h \rangle$ could be defined by using
the same approach as in the $\F_0 =1$ case, i.e. by using a
simplical path-integral representation to define them as
three-dimensional topologically invariant state-sums for the
$SU(2)$ quantum group at a root of unity. However, the main
problem in this approach will be the non-polynomiality of the $\O
(E)$ term, and hence one should resort to some sort of
approximation. The simplest thing one can do is to approximate
$\F_0$ by a coherent state. For example one can take \be \O(E) =
\int_\S d^3 x \, \sum_{i,a}(E_a^i - \d_a^i)^2
\quad,\label{polo}\ee so that $\F_0 = e^{-\O(E)}$ will be peaked
around the flat metric triads $e_i^a = \d_i^a$. However, the
expression (\ref{polo}) is not a diffeomorphism invariant. One
can try to replace it by a diffeomorphism invariant analog, for
example \be \O(E) = \int_\S d^3 x \, ( \det (E_a^i) -1 )^2
\quad.\label{ipolo}\ee However, (\ref{ipolo}) is not quite a
diffeomorphism invariant, since $\det (E)$ is a scalar density.
Still, given a polynomial $\O (E)$, one can use the path-integral
techniques developed for the simplical BF theories with interactions
\cite{fksf,mym} in order to obtain a state sum for a given
traingulation of $\S$. This state sum will not be topologically
invariant if the initial $\O(E)$ was not a
diffeomorphism invariant, and therefore some sort of averaging over
the topologically equivalent triangulations would have to be
performed.

Note that this approach suggests relaxing the condition that
$\Psi_0(A)$ should be annihilated by $\hat A -\hat \o$ to
requiring that $\Psi_0 (A)$ should be sufficiently peaked around
the flat metric triads, which means using a coherent state $\F_0$.
Then the expressions (\ref{flspi}) and (\ref{flsv}) also make
sense in the Dirac quantization formalism.

In the case when $\L \ne 0$, one would need to replace the Kodama
wavefunction $\Psi_K (A)$ by $\Psi_K (A) \Phi_0 (A)$ in order to
construct a flat metric vacuum because the de-Sitter classical
vacuum solution found in \cite{Smo} satisfies $F^a_{ij}-\L
\e_{ijk}E^{ka} =0$ and $R_3 =0$, where $R_3$ is the spatial
curvature, while the Kodama state corresponds only to $F^a_{ij}-\L
\e_{ijk}E^{ka}=0$ solutions without the flat metric condition.

Matter could be included by perturbing the vacuum state via the
ansatz $\Psi_0 (A) \to \Psi_0 (A)\chi [A,\phi]$, where $\f$ stands
for the matter degrees of freedom, see \cite{Smo}. In a similar
manner one can also study the gravitational perturbations, i.e.
the gravitons. Beside exploring the physical implications of the
presence of matter, it would be also interesting to see what kind
of invariants one would obtain by using the spin network
formulation and by defining the corresponding path-integrals via
generalized spin foam state sums. These can be then compared to
the generalized spin foam state sum amplitudes proposed for matter
in \cite{msf,mct,mym}.

\bigskip

\noindent{\bf ACKNOWLEDGEMENTS}

I would like to thank Nuno Costa-Dias and Roger Picken for the
discussions. Work supported by the FCT grants POCTI/FNU/
49543/2002 and POCTI/MAT/45306/2002.

\end{document}